\documentclass[12pt]{article}
\usepackage{latexsym, amssymb, amsmath,amsthm}

\def \F {{\mathbb F}}
\def \Z {{\mathbb Z}}
\def \N {{\mathbb N}}

\def \s {\sigma}

\newtheorem{theorem}{Theorem}
\newtheorem{definition}{Definition}

\newtheorem{remark}{Remark}

\newtheorem{corollary}{Corollary}

\title{Multisequences with high joint nonlinear complexity}

\author{Wilfried Meidl and Harald Niederreiter}

\date{\today}

\begin{document}

\maketitle

\begin{abstract}
We introduce the new concept of joint nonlinear complexity for multisequences over finite fields and we analyze the joint nonlinear complexity of 
two families of explicit inversive multisequences. We also establish a probabilistic result on the behavior of the joint nonlinear complexity of
random multisequences over a fixed finite field.
\end{abstract}

\section{Introduction} \label{sein}

There is a well-developed area that studies sequences over finite fields from the complexity-theoretic standpoint, with a view towards applications
in cryptography and pseudorandom number generation. We refer to the recent handbook article~\cite{mw13} for a concise survey of this area. For
applications that involve parallelization, such as word-based stream ciphers and pseudorandom vector generation, it is necessary to use multisequences
over finite fields. The complexity analysis of multisequences over finite fields has so far concentrated on the consideration of the joint linear
complexity and of closely related complexity measures for multisequences (see again~\cite{mw13}). In this paper, we introduce and analyze joint nonlinear
complexities of multisequences over finite fields.  

We use the standard notation $\F_q$ for the finite field with $q$ elements, where $q$ is a prime power. We abbreviate a sequence $\sigma_0,\sigma_1,\ldots$ of
elements of $\F_q$ by $(\sigma_i)_{i=0}^{\infty}$. For an integer $m \ge 1$, an $m$-\emph{fold multisequence over} $\F_q$ consists of $m$ parallel streams of
sequences of elements of $\F_q$. A multisequence may also be regarded as a sequence of vectors, and this viewpoint will be useful in Section~\ref{sepr}. 
Strictly speaking, for $m=1$ we have the case of a single sequence and not that of a multisequence in the usual sense, but we include the case $m=1$ for the
sake of completeness. A single sequence is therefore viewed as a $1$-fold multisequence. 

\begin{definition} \label{denc} {\rm
Let $\mathbf{Z} = (Z^{(1)},Z^{(2)},\ldots,Z^{(m)})$, $Z^{(j)} = (\sigma_i^{(j)})_{i=0}^\infty$, $1\le j \le m$, be an
$m$-fold multisequence over the finite field $\F_q$ and let $k,n \in \N$. The \emph{$n$th joint nonlinear complexity of order} $k$ of
the $m$-fold multisequence $\mathbf{Z}$, denoted by $N_k^{(m)}(\mathbf{Z},n)$, is the smallest $c \in \N$ for which there
exists a polynomial $f\in\F_q[x_1,\ldots,x_c]$ of degree at most $k$ in each variable such that
\begin{equation} \label{eqre} 
\s_{i+c}^{(j)} = f(\s_i^{(j)},\s_{i+1}^{(j)},\ldots,\s_{i+c-1}^{(j)}) \quad \mbox{for } 0\le i\le n-c-1,\;1\le j\le m. 
\end{equation}
This definition actually refers to the case where not all the first $n$ terms of all sequences $Z^{(j)}$, $1 \le j \le m$, are equal to $0$. Otherwise, we define
$N_k^{(m)}(\mathbf{Z},n)=0$.}
\end{definition}

\begin{remark} \label{renc} {\rm
We note that the definition of $N_k^{(m)}(\mathbf{Z},n)$ in Definition~\ref{denc} makes sense also if $\mathbf{Z}$ is a finite $m$-fold multisequence over $\F_q$
of length at least $n$. We always have $0 \le N_k^{(m)}(\mathbf{Z},n) \le n$. }
\end{remark}

\begin{remark} \label{remo} {\rm
In Definition~\ref{denc} it suffices to consider the case where $1 \le k \le q-1$. This follows from the well-known fact that, as a map, any polynomial $f: \F_q^c
\to \F_q$ can be represented by a polynomial over $\F_q$ in $c$ variables of degree at most $q-1$ in each variable (see \cite[pp. 368--369]{ln}). Thus, for $k \ge q-1$
all joint nonlinear complexities $N_k^{(m)}(\mathbf{Z},n)$ of a fixed $\mathbf{Z}$ are the same and equal to $N_{q-1}^{(m)}(\mathbf{Z},n)$. For $k=q-1$ and $m=1$,
$N_{q-1}^{(1)}(\mathbf{Z},n)$ is equal to the \emph{$n$th maximum-order complexity} introduced by Jansen~\cite{ja} and studied further in~\cite{em}, \cite{ja91}, \cite{jb}, \cite{lkk},
\cite{lkk07}, and~\cite{hc}. For arbitrary $m \ge 1$, it is reasonable to call $N_{q-1}^{(m)}(\mathbf{Z},n)$ the \emph{$n$th joint maximum-order complexity} of $\mathbf{Z}$.
The definition in~\cite[Definition~1]{ja91} may be viewed as a previous notion of joint maximum-order complexity.}
\end{remark}
 
We apply the joint nonlinear complexities in Definition~\ref{denc} to the analysis of the well-known family of explicit inversive pseudorandom sequences. We show that,
by combining suitably chosen explicit inversive pseudorandom sequences into multisequences, we can construct multisequences with high joint nonlinear complexities.
Sections~\ref{seeq} and~\ref{seet} contain appropriate constructions and results for two different types of explicit inversive pseudorandom sequences. In Section~\ref{sepr},
we establish a benchmark result on the behavior of joint nonlinear complexities of random multisequences over $\F_q$.

\section{Explicit inversive pseudorandom number generator with period $q$} \label{seeq}

For a prime $p$ and a positive integer $r$, let $q = p^r$. We identify the finite prime field $\F_p$ with the set $\Z_p := \{0,1,\ldots,p-1\} \subset \Z$ with
arithmetic modulo $p$. For a fixed basis $\{\gamma_1,\gamma_2,\ldots,\gamma_r\}$ of $\F_q$ over $\F_p$, we define 
$$ 
\xi_n = n_1\gamma_1+n_2\gamma_2 + \cdots + n_r\gamma_r \quad \mbox{for } n=0,1,\ldots,q-1 
$$
if 
$$ 
n = n_1+n_2p + \cdots + n_rp^{r-1} \quad \mbox{with } n_1,n_2,\ldots,n_r \in \Z_p.
$$
We extend the definition of the $\xi_n$ periodically with period $q$, so that we have $\xi_{n+q} = \xi_n$ for all $n\ge 0$.

We define an $m$-fold multisequence $\mathbf{S}_r = (S^{(1)},S^{(2)},\ldots,S^{(m)})$, 
$S^{(j)} = (\sigma_i^{(j)})_{i=0}^\infty$, $1\le j \le m$, over $\F_q$ by choosing $\beta_1,\ldots,\beta_m \in \F_q^*$ and putting
\begin{equation} \label{Sr}
\sigma_i^{(j)} = (\xi_i+\beta_j)^{q-2}\quad \mbox{for } i\ge 0 \mbox{ and } 1 \le j \le m. 
\end{equation}
Note that $\mathbf{S}_r$ is periodic with least period $q$.

For $m=1$ and $r=1$ this generator was introduced in~\cite{ei} and for $m=1$ and arbitrary $r$ in~\cite{ha}.
The linear complexity profile of this single sequence has been analyzed in~\cite{mw03} (see~\cite{mw13} for the definition of
the linear complexity profile).
In~\cite{mw03}, again for $m=1$, the nonlinear complexity profile (see~\cite{mw03} and~\cite{hc}), which is a less general concept than the concept
in Definition~\ref{denc}, has been investigated for the explicit inversive pseudorandom number generator~$(\ref{Sr})$. 
The nonlinear complexity profile for some recursively defined generators has been estimated in~\cite{gsw}. 
For $m$ greater than $1$, the generator~$(\ref{Sr})$ was first considered in~\cite{mw}, where bounds on its joint linear complexity 
profile have been established (see again~\cite{mw13} for the definition of the joint linear complexity profile).

The objective in this section is to analyze the $n$th joint nonlinear complexity of order $k$ of the $m$-fold multisequence~$(\ref{Sr})$. 
To the best of our knowledge, this is the first treatment of the joint nonlinear complexity of a concrete multisequence generator.

We put
$$
W(a)=\min (a,p-a) \quad \mbox{for any } a \in \Z_p.
$$
For any $\alpha \in \F_q$, let
$$
\alpha =a_1 \gamma_1 + a_2 \gamma_2 + \cdots + a_r \gamma_r \quad \mbox{with } a_1,a_2,\ldots,a_r \in \Z_p
$$
be the unique representation of $\alpha$ as a linear combination of the basis elements $\gamma_1,\gamma_2,\ldots,\gamma_r$. Then we define
$$
||\alpha||=\sum_{s=1}^r W(a_s) p^{s-1}.
$$
For later use, we note that $||\alpha||=||-\alpha||$ for all $\alpha \in \F_q$.

Let $m$ be an integer with $1 \le m \le q-1$. We choose pairwise distinct elements $\beta_1,\ldots,\beta_m \in\F_q^*$. 
For $m \ge 2$ we define the minimum distance $d_r$ between $\beta_1,\ldots,\beta_m$ as
$$ 
d_r = d_r(\beta_1,\ldots,\beta_m)= \min_{1\le j_1 <j_2 \le m} \; ||\beta_{j_1}-\beta_{j_2}||\quad \mbox{for } m \ge 2. 
$$
For $m=1$, by convention $d_r := q$. Note that we always have $1 \le d_r \le q$.
We also remark that our definition of $d_r$ is a corrected version of the definition in \cite{mw}.
The results in \cite{mw} on the joint linear complexity profile hold with our definition of $d_r$.
 
\begin{theorem} \label{bound0}
Let $\mathbf{S}_r$ be an $m$-fold multisequence over $\F_q$ of the form~\eqref{Sr} with $1 \le m \le q-1$ and pairwise distinct $\beta_1,\ldots,\beta_m \in \F_q^*$. 
Then for integers $1\le k\le q-1$ and $1 \le n \le q-1$, the $n$th joint nonlinear complexity of order $k$ of $\mathbf{S}_r$ satisfies
$$ 
N_k^{(m)}(\mathbf{S}_r,n) \ge \min\left(\frac{n}{2},\sqrt{\frac{mn}{4(k+3)}},d_r,\sqrt{\frac{m(d_r+1)}{4(k+3)}}\right). 
$$
\end{theorem}

{\it Proof.}
We fix $k$, $m$, and $n$, and so we may use the abbreviated notation $N_k^{(m)}(\mathbf{S}_r,n) = c_n = c$. 
Note that $c > 0$ since $\sigma_0^{(j)} = \beta_j^{-1}$, hence every component sequence $S^{(j)}$, $1\le j\le m$,
starts with a nonzero element of $\F_q$. 

First we consider the case where $n < c+d_r+1$.
Suppose that $f \in \F_q[x_1,\ldots,x_c]$, $1\le c \le n-1$, is a polynomial of degree at most $k$ in each variable
such that
$$ 
\s_{i+c}^{(j)} = f(\s_i^{(j)},\s_{i+1}^{(j)},\ldots,\s_{i+c-1}^{(j)}) \quad \mbox{for } 0\le i\le n-c-1,\;1\le j\le m. 
$$
Then for those integers $0\le i\le n-c-1$, $1\le j\le m$, for which $\xi_{i+l}+\beta_j \ne 0$ for all $l = 0,1,\ldots,c$, we have
\begin{equation} \label{p1} 
-\frac{1}{\xi_{i+c}+\beta_j} + f\left(\frac{1}{\xi_i+\beta_j},\frac{1}{\xi_{i+1}+\beta_j},\ldots,\frac{1}{\xi_{i+c-1}+\beta_j}\right) = 0. 
\end{equation}
We exclusively consider those integers $i$, $0\le i\le n-c-1$, for which we additionally have
$\xi_{i+l} = \xi_i + \xi_l$ for all $0\le l\le c$. Then~\eqref{p1} is equivalent to
\begin{equation} \label{p2} 
-\frac{1}{\xi_i+\xi_c+\beta_j} + f\left(\frac{1}{\xi_i+\beta_j},\frac{1}{\xi_i+\xi_1+\beta_j},\ldots,\frac{1}{\xi_i+\xi_{c-1}+\beta_j}\right) = 0. 
\end{equation}
Consequently, all elements of the form
\begin{equation} \label{roots}
\lambda = \xi_i+\beta_j \quad \mbox{for } 0\le i \le n-c-1, \; 1\le j\le m, 
\end{equation}
such that
$$
\xi_{i+l} = \xi_i + \xi_l \quad \mbox{and} \quad \lambda + \xi_l \ne 0 \quad \mbox{for } 0\le l\le c 
$$
are zeros of the rational function
\begin{equation} \label{R}
R(z) = -\frac{1}{z+\xi_c} + f\left(\frac{1}{z},\frac{1}{z+\xi_1},\ldots,\frac{1}{z+\xi_{c-1}}\right). 
\end{equation}
We may suppose that $c < d_r$ (and thus $c<q$), for otherwise the lower bound in the theorem holds trivially. Then $-\xi_c$ is not a pole of 
$f\left(\frac{1}{z},\frac{1}{z+\xi_1},\ldots,\frac{1}{z+\xi_{c-1}}\right)$,
hence $R(z)= g(z)/h(z) \neq 0 \in \F_q(z)$. If $R = g/h$ is reduced to lowest degree terms, then by the definition of $R$
the polynomials $g,h \in \F_q[z]$ satisfy $\deg(g) \le \deg(h) \le kc+1$.

To estimate the number of elements of the form~\eqref{roots}, we define integers $0 \le v < r$, $0 \le w < r$, $1 \le N_v < p$, and $1 \le L_w < p$ by
$$ 
N_vp^v \le n < (N_v+1)p^v \quad \mbox{and} \quad L_wp^w \le c < (L_w+1)p^w. 
$$
Since $c \le n$, we have $w < v$ or $w = v$ and $L_v\le N_v$.

First suppose that $w<v$. Then (compare with~\cite[Section~3]{mw}) we have $\xi_{i+l} = \xi_i + \xi_l$, $0\le l\le c$, for $\xi_i$ with
$i = h_wp^w + \cdots + h_vp^v$, where $h_w,\ldots,h_v \in \Z_p$, $0\le h_w < p-L_w$, and $0\le h_v < N_v$.
Note that $i+l \le i+c \le n-1$. Consequently, we have at least
\begin{equation} \label{minsol} 
N_v(p-L_w)p^{v-w-1} > \frac{N_v}{N_v+1}\frac{(p-L_w)L_wn}{pc} \ge \frac{n}{4c} 
\end{equation}
distinct elements $\xi_i$ satisfying $\xi_{i+l} = \xi_i + \xi_l$, $0\le l\le c$.

We show next that the elements $\xi_i+\beta_j$ in~\eqref{roots}, with $i$ as in the preceding paragraph, are all distinct if $n < c+d_r+1$. 
So suppose that $\xi_{i_1}+\beta_{j_1} = \xi_{i_2}+\beta_{j_2}$ with $1 \le j_1,j_2 \le m$ and $j_1 \ne j_2$. Then 
$$ 
\xi_{i_1}-\xi_{i_2} = \beta_{j_2}-\beta_{j_1} = b_w\gamma_{w+1}+b_{w+1}\gamma_{w+2} + \cdots + b_v\gamma_{v+1} 
$$
with $b_w,b_{w+1},\ldots,b_v \in \Z_p$. We may assume that $0 \le b_v < N_v$, otherwise we consider the equality $\xi_{i_2}-\xi_{i_1} = \beta_{j_1}-\beta_{j_2}$.
By the definitions of $||\beta_{j_1}-\beta_{j_2}||$ and $d_r$, we have 
$$ 
d_r \le ||\beta_{j_1}-\beta_{j_2}|| = e_wp^w+e_{w+1}p^{w+1} + \cdots + e_vp^v, 
$$
where $e_w,e_{w+1},\ldots,e_v \in \Z_p$, $0 \le e_w \le p-L_w-1$, and $0 \le e_v \le N_v-1$. Consequently,
\begin{eqnarray*}
 n-c & > & N_vp^v - (L_w+1)p^w \\
& = & (N_v-1)p^v + \sum_{s=w+1}^{v-1} (p-1)p^s +(p-L_w-1)p^w \\
& \ge & \sum_{s=w}^v e_sp^s = ||\beta_{j_1}-\beta_{j_2}|| \ge d_r, 
\end{eqnarray*}
which contradicts $n < c+d_r+1$.

As a consequence, the rational function $R$ has at least $m\frac{n}{4c}-(c+1)$ distinct zeros.
Therefore, together with the previous upper bound on the degree of the numerator $g$ of $R$, we obtain $kc+1 \ge \frac{mn}{4c}-(c+1)$, 
thus $c(k+3) \ge c(k+1)+2 \ge \frac{mn}{4c}$, or equivalently
$$ 
c \ge \sqrt{\frac{mn}{4(k+3)}}. 
$$

Secondly, we investigate the case where $w=v$. If $L_v = N_v$, then $c \ge N_vp^v > (N_v/(N_v+1))n \ge n/2$.
Now let $w=v$ and $N_v \ge L_v+1 \ge 2$. Then we have $\xi_{i+l} = \xi_i + \xi_l$ for $0\le l \le c$ for
at least the $(N_v-L_v)$ distinct elements $\xi_i$ with $i = h_vp^v$ and $0\le h_v \le N_v-L_v-1$. As before, we want to show that the elements $\xi_i+\beta_j$
in~\eqref{roots} are distinct if $n <c+d_r+1$, where now $i=h_vp^v$ and $0 \le h_v \le N_v-L_v-1$. Note that then $\xi_i=h_v\gamma_{v+1}$, so if we had
$h_v\gamma_{v+1}+\beta_{j_1}=h_v^{\prime}\gamma_{v+1}+\beta_{j_2}$ with $0 \le h_v < h_v^{\prime} \le N_v-L_v-1$, then
$$
d_r \le ||\beta_{j_1}-\beta_{j_2}|| = ||(h_v^{\prime}-h_v)\gamma_{v+1}|| \le (N_v-L_v-1)p^v < n-c,
$$
which is a contradiction.

It follows that the rational function $R$ has at least $m(N_v-L_v) - c-1$ zeros, hence
\begin{equation} \label{w=v}
c(k+3) \ge c(k+1) + 2 \ge m(N_v-L_v).
\end{equation}
Suppose that
$$ 
m > 3\left(\frac{L_v}{N_v+1}\right)^2(k+3)n. 
$$
Then~\eqref{w=v} yields
$$ 
c \ge \frac{m(N_v-L_v)}{k+3} > \sqrt{mn}\frac{(N_v-L_v)L_v\sqrt{3}}{(N_v+1)\sqrt{k+3}} \ge \sqrt{\frac{mn}{3(k+3)}}. 
$$
Otherwise, we have
$$ 
\left(\frac{L_v}{N_v+1}\right)^2 \ge \frac{m}{3n(k+3)} 
$$
and we again obtain
$$ 
c \ge L_vp^v > \frac{L_v}{N_v+1}n \ge \sqrt{\frac{mn}{3(k+3)}}. 
$$
Altogether, assuming that $n < c+d_r+1$, we have
$$ 
c \ge \min\left(\frac{n}{2},\sqrt{\frac{mn}{4(k+3)}}\right). 
$$

If $n \ge c_n+d_r+1=c+d_r+1$, then $c \ge c_{d_r+c}$. Hence by what we have already shown, we have either $c \ge \frac{d_r+c}{2}$, i.e., $c \ge d_r$, or
$c \ge \sqrt{\frac{m(d_r+c)}{4(k+3)}} \ge \sqrt{\frac{m(d_r+1)}{4(k+3)}}$. Now the desired lower bound on $c$ is proved in all cases. \hfill$\Box$

\section{Explicit inversive pseudorandom number generator with period $t$} \label{seet}

Let $\F_q$ be the finite field with $q \ge 3$ elements, let $t$ be a positive divisor of $q-1$, and let $\gamma \in \F_q^*$ be an
element of order $t$. Let $m$ be an integer with $2 \le m \le q-1$. We choose pairwise distinct elements $\alpha_1,\ldots, \alpha_m \in \F_q^*$ and an element 
$\beta\in\F_q^*$. Then we define an $m$-fold multisequence $\mathbf{Z} = (Z^{(1)},Z^{(2)},\ldots,Z^{(m)})$, $Z^{(j)} = (\sigma_i^{(j)})_{i=0}^\infty$, 
$1\le j \le m$, over $\F_q$ by
\begin{equation} \label{Z}
\sigma_i^{(j)} = (\alpha_j\gamma^i-\beta)^{q-2} \quad \mbox{for } i\ge 0 \mbox{ and } 1 \le j \le m. 
\end{equation} 
Note that $\mathbf{Z}$ is periodic with least period $t$.

The generator~$(\ref{Z})$ has been introduced in~\cite{mw04} and its linear complexity profile has been investigated. 
In~\cite[Section~4]{mw} the $n$th joint linear complexity $L^{(m)}_n(\mathbf{Z})$ of the multisequence~$(\ref{Z})$ has been analyzed.
In particular, it has been shown that some of those multisequences satisfy $L^{(m)}_n(\mathbf{Z}) \ge mn/(m+1)$
for small values of $n$, i.e. they exhibit a perfect joint linear complexity profile (cf.~\cite{x}).
Lower bounds on the $n$th nonlinear complexity of order $k$, defined as in Definition~\ref{denc} for $m=1$,
of the generator~$(\ref{Z})$ were obtained in~\cite{hc}. The results slightly improve bounds for the generators in \cite{gsw,mw03}.
In this section, an analysis of the $n$th joint nonlinear complexity of order $k$ of the generator~$(\ref{Z})$ is given for arbitrary 
values of $m\ge 2$. The results suggest that the multisequences~$(\ref{Z})$ also possess an excellent behavior with respect to 
joint nonlinear complexity.

For $\xi \in \F_q^*$ we define
$$ 
||\xi||_t = b \quad \mbox{if } \xi = \gamma^b \mbox{ with } 0 \le b < t 
$$
and $||\xi||_t = t$ if $\xi$ does not belong to the cyclic subgroup $\langle\gamma\rangle$ of $\F_q^*$ generated by $\gamma$. Furthermore, we define
$$ 
\delta_t =\delta_t(\alpha_1,\ldots,\alpha_m)= \min_{1 \le j_1, j_2 \le m \atop j_1 \ne j_2} \, ||\alpha_{j_1} \alpha_{j_2}^{-1}||_t. 
$$
Note that we always have $1 \le \delta_t \le t$.

\begin{theorem} \label{thet}
Let $\gamma \in \F_q^*$ with $q \ge 3$ be an element of order $t$ and let $\mathbf{Z}$ be an $m$-fold multisequence over $\F_q$ of the form~\eqref{Z} with $2 \le m \le q-1$.
Then for integers $1\le k\le q-1$ and $n \ge 1$, the $n$th joint nonlinear complexity of order $k$ of $\mathbf{Z}$ satisfies
$$ 
N_k^{(m)}(\mathbf{Z},n) \ge \min\left(\frac{mn-2}{m+k+1},\frac{\delta_tm-2}{k+1},t\right). 
$$
\end{theorem}

{\it Proof.}
Since the joint nonlinear complexity of order $k$ is invariant under the termwise multiplication of all component 
sequences with a fixed element from $\F_q^*$, we may assume that $\beta = 1$. We fix $k$, $m$, and $n$, and we write $c=N_k^{(m)}(\mathbf{Z},n)$. We have $c > 0$
since $\sigma_0^{(j)}=(\alpha_j-1)^{q-2} \ne 0$ for at least one $j$ with $1 \le j \le m$.

Suppose that $f \in \F_q[x_1,\ldots,x_c]$, $1\le c \le n-1$, is a polynomial of degree at most $k$ in each variable
such that
$$ 
\s_{i+c}^{(j)} = f(\s_i^{(j)},\s_{i+1}^{(j)},\ldots,\s_{i+c-1}^{(j)}) \quad \mbox{for } 0\le i\le n-c-1,\;1\le j\le m. 
$$
Consequently, for those integers $0 \le i \le n-c-1$ and indices $j$ for which $\alpha_j\gamma^{i+l} \ne 1$ for $0\le l\le c$, we have
$$ 
-\frac{1}{\alpha_j\gamma^{i+c}-1} + f\left(\frac{1}{\alpha_j\gamma^i-1},\frac{1}{\alpha_j\gamma^{i+1}-1},\ldots,\frac{1}{\alpha_j\gamma^{i+c-1}-1}\right) = 0. 
$$
Hence all elements of the form
$$
\lambda = \alpha_j\gamma^i \quad \mbox{for } 0\le i \le n-c-1,\;1\le j\le m,
$$
such that
$$
\alpha_j\gamma^{i+l} \ne 1 \quad \mbox{for } 0\le l\le c 
$$
are zeros of the rational function
\begin{eqnarray*} 
R(z) & = & -\frac{1}{\gamma^cz-1} + f\left(\frac{1}{z-1},\frac{1}{\gamma z-1},\ldots,\frac{1}{\gamma^{c-1}z-1}\right) \\
& = & -\frac{1}{\gamma^cz-1} + \frac{G(z)}{H(z)}. 
\end{eqnarray*}
We can suppose that $c < t$. Then the element $\gamma^{-c}$ is not a root of $H(z)$ and consequently not a root of 
$H(z) - (\gamma^cz-1)G(z)$. Hence $R(z) \ne 0 \in \F_q(z)$.

Write $R$ in reduced form as $g/h$ with $g,h \in \F_q[z]$ and $\gcd(g,h) = 1$. By the definition of $R$ we have
$\deg(g)\le\deg(h) \le kc+1$. If $n \le \delta_t + c$, then all $m(n-c)$ elements $\lambda = \alpha_j\gamma^i$, $0\le i \le n-c-1$, $1\le j\le m$, 
are distinct. Hence $g$ has at least $m(n-c)-(c+1)$ roots. It follows that 
$$
m(n-c)-(c+1) \le \deg(g) \le kc+1,
$$
and so $c \ge (mn-2)/(k+m+1)$.
If $n > \delta_t+c$, then all $m\delta_t$ elements $\alpha_j \gamma^i$, $0 \le i < \delta_t$, $1 \le j \le m$, are distinct, and so $g$ has at 
least $m\delta_t-(c+1)$ roots.
It follows that $c \ge (m\delta_t-2)/(k+1)$.
\hfill$\Box$\\[.5em]
 
For the case where $t < q-1$, the bound in Theorem~\ref{thet} can be improved if all $\alpha_j$,
$1 \le j \le m$, are chosen not to be in the coset $\beta \langle\gamma\rangle$ of $\langle\gamma\rangle$.

\begin{corollary}
Let $\gamma \in \F_q^*$ with $q \ge 3$ be an element of order $t < q-1$, let $\alpha_j \in \F_q^*\setminus \beta \langle\gamma\rangle$
for $1\le j\le m$, and let $\mathbf{Z}$ be an $m$-fold multisequence over $\F_q$ 
of the form~\eqref{Z} with $2 \le m \le q-1-t$. Then for integers $1\le k\le q-1$ and $n \ge 1$, the $n$th joint nonlinear complexity of order $k$ 
of $\mathbf{Z}$ satisfies
$$ 
N_k^{(m)}(\mathbf{Z},n) \ge \min\left(\frac{mn-1}{m+k},\frac{\delta_tm-1}{k},t\right).
$$
\end{corollary}

{\it Proof.}
As in the proof of Theorem~\ref{thet}, we can assume that $\beta =1$.
Since we then suppose that $\alpha_j\gamma^r \ne 1$ for all $1 \le j \le m$ and for all integers $r$, we need not subtract $c+1$ 
when estimating the number of roots of $g$ in the proof of Theorem~\ref{thet}.
Consequently, if $n \le \delta_t + c$, then $g$ has at least $m(n-c)$ roots, and so $c \ge (mn-1)/(k+m)$.
If $n > \delta_t+c$, then $g$ has at least $m\delta_t$ roots, and hence $c \ge (m\delta_t-1)/k$.
\hfill$\Box$

\section{A probabilistic result} \label{sepr}

We establish a probabilistic result on the behavior of joint nonlinear complexities of random multisequences over the finite field $\F_q$, where $q$
is an arbitrary prime power.  For a positive integer $m$, the set of all $m$-fold multisequences over $\F_q$ can be identified with the set $(\F_q^m)^{\infty}$
of all sequences over $\F_q^m$. In other words, $(\F_q^m)^{\infty}$ is the Cartesian product of denumerably many copies of $\F_q^m$. We introduce a
canonical probability measure on $(\F_q^m)^{\infty}$ as follows. Let $\mu_{q,m}$ be the uniform probability measure on $\F_q^m$ which assigns the measure $q^{-m}$ to 
each element of $\F_q^m$. Then $\mu_{q,m}^{\infty}$ is the complete product probability measure on $(\F_q^m)^{\infty}$ induced by $\mu_{q,m}$. 

We say that a property of $m$-fold multisequences $\mathbf{Z} \in (\F_q^m)^{\infty}$ holds $\mu_{q,m}^{\infty}$-\emph{almost everywhere} if it holds for a set 
of $m$-fold multisequences $\mathbf{Z}$ of $\mu_{q,m}^{\infty}$-measure $1$.
We may view such a property as a typical property of a random $m$-fold multisequence over $\F_q$.

\begin{theorem} \label{thpr}
Let $k$ and $m$ be integers with $1 \le k \le q-1$ and $m \ge 1$. Then $\mu_{q,m}^{\infty}$-almost everywhere we have
$$
\liminf_{n \to \infty} \Big(N_k^{(m)}(\mathbf{Z},n)-\frac{\log (mn)}{\log (k+1)} \Big) \ge 0.
$$
\end{theorem}

{\it Proof.}
We fix $k$, $m$, and $q$ throughout the proof.
For $n,r \in \N$ with $r \le n$, let $T_{k,n}^{(m)}(r)$ be the number of $m$-fold multisequences $\mathbf{Z}_n$ over $\F_q$ of length $n$ with $N_k^{(m)}(\mathbf{Z}_n,n) \le r$. 
We view each $\mathbf{Z}_n$ as a finite sequence $\mathbf{Z}_n=(\mathbf{s}_i)_{i=0}^{n-1}$ of vectors $\mathbf{s}_i \in \F_q^m$. Each sequence 
$\mathbf{Z}_n =(\mathbf{s}_i)_{i=0}^{n-1}$ counted by $T_{k,n}^{(m)}(r)$ is (not necessarily uniquely) determined by a polynomial $f \in \F_q[x_1,\ldots,x_r]$ of degree at most
$k$ in each variable and by initial vectors $\mathbf{s}_0,\mathbf{s}_1,\ldots,\mathbf{s}_{r-1} \in \F_q^m$ in a vector recursion
$$
\mathbf{s}_{i+r}=f(\mathbf{s}_i,\mathbf{s}_{i+1},\ldots,\mathbf{s}_{i+r-1}) \quad \mbox{for } 0 \le i \le n-r-1.
$$
Here we have written a recursion of the form~\eqref{eqre} in vector notation in an obvious manner, i.e., the polynomial $f$ operates on each of the $m$ components of the
vectors $\mathbf{s}_i,\mathbf{s}_{i+1},\ldots,\mathbf{s}_{i+r-1}$. The number of possibilities for $f$ is $q^{(k+1)^r}$ and the number of
possibilities for the $r$ initial vectors from $\F_q^m$ is $q^{mr}$. Therefore
\begin{equation} \label{eqtk}
T_{k,n}^{(m)}(r) \le q^{(k+1)^r +mr} \quad \mbox{for } 1 \le r \le n.
\end{equation}
Now we fix $\varepsilon > 0$ and put
$$
B_n=\frac{\log (mn)}{\log (k+1)} -\varepsilon \quad \mbox{for } n=1,2,\ldots
$$
and
$$
\mathcal{A}_n=\{\mathbf{Z} \in (\F_q^m)^{\infty} : N_k^{(m)}(\mathbf{Z},n) \le B_n \} \quad \mbox{for } n=1,2,\ldots .
$$
We have $1 \le \lfloor B_n \rfloor \le n$ for sufficiently large $n$. Since $N_k^{(m)}(\mathbf{Z},n)$ depends only on the first $n$ terms of $\mathbf{Z}$, the bound~\eqref{eqtk} yields
\begin{equation} \label{eqan}
\mu_{q,m}^{\infty}(\mathcal{A}_n) = q^{-mn} T_{k,n}^{(m)}(\lfloor B_n \rfloor ) \le q^{(k+1)^{B_n} +mB_n-mn}
\end{equation}
for sufficiently large $n$. The definition of $B_n$ implies that
$$
(k+1)^{B_n} +mB_n-mn < n \Big(\frac{m}{(k+1)^{\varepsilon}} + \frac{m \log (mn)}{n \log(k+1)} -m \Big). 
$$
Now
$$
\lim_{n \to \infty} \Big(\frac{m}{(k+1)^{\varepsilon}} + \frac{m \log (mn)}{n \log (k+1)} -m \Big) = \frac{m}{(k+1)^{\varepsilon}} - m < 0,
$$
and so for some $0 < \delta < 1$ we have
$$
(k+1)^{B_n} +mB_n -mn < -\delta n
$$
for sufficiently large $n$. It follows then from~\eqref{eqan} that $\sum_{n=1}^{\infty} \mu_{q,m}^{\infty}(\mathcal{A}_n) < \infty$. Then the Borel-Cantelli lemma 
(see \cite[Lemma~3.14]{br} and \cite[p.~228]{lo}) shows that the set of all $\mathbf{Z} \in (\F_q^m)^{\infty}$ for which $\mathbf{Z} \in \mathcal{A}_n$ for infinitely 
many $n$ has $\mu_{q,m}^{\infty}$-measure $0$. In other words, $\mu_{q,m}^{\infty}$-almost everywhere we have $\mathbf{Z} \in \mathcal{A}_n$ for at most finitely many $n$. 
By the definition of $\mathcal{A}_n$, this means that $\mu_{q,m}^{\infty}$-almost everywhere the inequality
$$
N_k^{(m)}(\mathbf{Z},n) > B_n =\frac{\log (mn)}{\log (k+1)} - \varepsilon
$$
is satisfied for sufficiently large $n$. Consequently, $\mu_{q,m}^{\infty}$-almost everywhere we have 
$$
\liminf_{n \to \infty} \Big(N_k^{(m)}(\mathbf{Z},n) - \frac{\log (mn)}{\log (k+1)} \Big) \ge - \varepsilon.
$$
By applying this for all $\varepsilon =1/l$ with $l \in \N$ and noting that the intersection of countably many sets of $\mu_{q,m}^{\infty}$-measure $1$ has
again $\mu_{q,m}^{\infty}$-measure $1$, we obtain the result of the theorem.
\hfill$\Box$

\begin{remark} \label{repr} {\rm
For $k=q-1$ and $m=1$, that is, for the maximum-order complexity (see Remark~\ref{remo}), results of Jansen~\cite{ja} and Erdmann and Murphy~\cite{em} demonstrate 
that the expected value of
$N_{q-1}^{(1)}(\mathbf{Z},n)$ behaves asymptotically like $(\log n)/(\log q)$, up to an absolute constant. On the basis of these results and of Theorem~\ref{thpr}, 
we venture the conjecture that for any $m \ge 1$ we have
$$
\lim_{n \to \infty} \frac{N_{q-1}^{(m)}(\mathbf{Z},n)}{\log (mn)} = C_q^{(m)} \quad \mu_{q,m}^{\infty}\mbox{-almost everywhere},
$$
where the constant $C_q^{(m)} > 0$ depends only on $q$ and $m$. In view of the heuristic that the expected order of magnitude of $N_k^{(m)}(\mathbf{Z},n)$ for random
$m$-fold multisequences $\mathbf{Z}$ over $\F_q$ is $\log (mn)$, it is clear that the multisequences in Sections~\ref{seeq} and~\ref{seet} can be said to have high joint
nonlinear complexity under suitable conditions on their parameters. }
\end{remark}

\vspace{1cm}

\noindent
Wilfried Meidl, Sabanci University, MDBF, Tuzla, 34956 Istanbul, Turkey; \\
email: meidlwilfried@gmail.com \\

\noindent
Harald Niederreiter, Johann Radon Institute for Computational and Applied Mathematics, Austrian Academy of Sciences, Altenbergerstr. 69, A-4040 Linz, Austria,
and Department of Mathematics, University of Salzburg, Hellbrunnerstr. 34, A-5020 Salzburg, Austria; email: ghnied@gmail.com

\end{document}